\documentclass[10pt]{iopart}

\usepackage{pgfplots}
\usepgfplotslibrary{colormaps}
\usepackage{pgfplotstable}
\pgfplotsset{compat=newest}
\usetikzlibrary{calc}

\usepackage{graphicx}
\usepackage{colortbl}
\usepackage[ruled]{algorithm2e}
\usepackage{setspace}
\usepackage{csquotes}
\MakeOuterQuote{"}
\usepackage{todonotes}
\usepackage{amsmath}
\usepackage{amssymb}
\usepackage{relsize}
\usepackage{parskip}

\usepackage{pifont}
\DeclareRobustCommand{\textbigstar}{\mbox{\ding{72}}}

\usepackage{geometry}
\usepackage{xcolor}
\usepackage{soul}
\sethlcolor{yellow}
\usepackage{siunitx}

\usepackage[labelfont=bf]{caption}

\usepackage{natbib}

\usepackage{hyperref}
\hypersetup{colorlinks=true, linkcolor=black, urlcolor=blue, citecolor=blue}

\begin{document}

\title[]{Dimensionality reduction of neuronal degeneracy reveals two interfering physiological mechanisms} 

\author{Arthur Fyon$^{1, *}$, Alessio Franci$^{1,2}$, Pierre Sacré$^1$, and Guillaume Drion$^{1, 3, *}$}

\address{%
    $^1$ University of Liège, Department of Electrical Engineering and Computer Science, B-4000 Liège, Belgium.\\
    $^2$ WEL Research Institute, B-1300 Wavre, Belgium. \\
    $^3$ Lead contact.\\
    $^{*}$ Correspondence: \href{mailto:afyon@uliege.be}{afyon@uliege.be} (A.Fy.), \href{mailto:gdrion@uliege.be}{gdrion@uliege.be} (G.D.).
    }

\begin{figure}[ht!]
\centering
\includegraphics[width=0.9\linewidth]{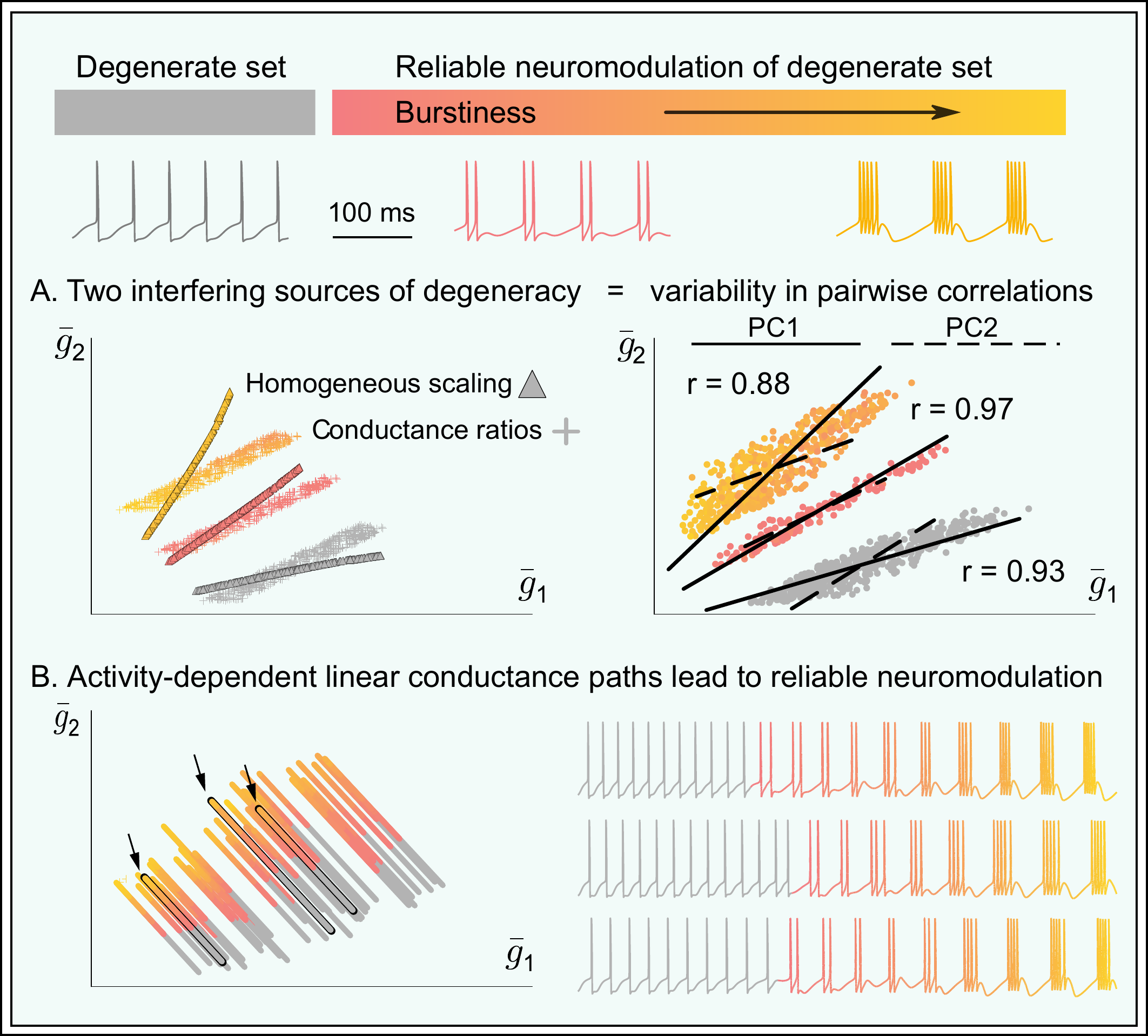}
\end{figure}

\noindent{\it Ion channel correlation, neuronal variability, neuronal degeneracy, conductance-based models, neuromodulation}
\maketitle

\section*{\textbf{SUMMARY}}
Neuronal systems maintain stable functions despite large variability in their physiological components. Ion channel expression, in particular, is highly variable in neurons exhibiting similar electrophysiological phenotypes, which poses questions regarding how specific ion channel subsets reliably shape neuron intrinsic properties. Here, we use detailed conductance-based modeling to explore the origin of stable neuronal function from variable channel composition. Using dimensionality reduction, we uncover two principal dimensions in the channel conductance space that capture most of the variance of the observed variability. Those two dimensions correspond to two physiologically relevant sources of variability that can be explained by feedback mechanisms underlying regulation of neuronal activity, providing quantitative insights into how channel composition links to neuronal electrophysiological activity. These insights allowed us to understand and design a model-independent, reliable neuromodulation rule for variable neuronal populations.

\section*{\textbf{INTRODUCTION}}
A remarkable property of nervous systems is their ability to maintain stable functions despite large variability and turnover of the underlying physiological components. This observation has led to the understanding that neuron electrophysiological properties are shaped by the coordinated expression of potentially large subsets of ion channels \citep{goaillard2021ion}, which represent a substantial challenge in any attempt to link ion channel properties with neuron electrophysiological signature. 

Over the last decades, a combination of experimental and computational work has provided insights into the relationship between ion channel densities and neuronal signaling. 
First, it is now clear that different combinations of ion channels can lead to a similar activity from highly variable channel densities \citep{prinz2004similar, achard2006complex, alonso2019visualization, taylor2009multiple, swensen2005robustness}, due to a functional overlap in channel voltage- and time-dependent properties \citep{goaillard2021ion, drion2015dynamic}. 
Second, it has been experimentally shown that the ion channel expression correlate positively in a same neuron type, and that different neuron types show different correlation graphs \citep{schulz2006variable, schulz2007quantitative, amendola2012ca2, liss2001tuning, schultz2007multiple, tobin2009correlations}. Such positive correlations in ion channel expression have been shown to emerge from physiologically plausible homeostatic rules \citep{o2014cell}. One could therefore argue that specific correlation graphs in channel expressions are an important neuronal signature. 
Third, reliable neuromodulation has been shown to often occur through a concomitant action on several channel subtypes \citep{amendola2012ca2, nadim2014neuromodulation, grashow2009reliable, schulz2006cellular, marder2007understanding}, which highlights the importance of understanding the mechanisms linking ion channel density and neuronal signaling. 

Although this body of work has deepened our understanding of how ion channels shape neuronal activity, many important questions remain open. 
First, although positive correlations in ion channel expression have largely been reported, studies on correlation in actual conductance values show a blurrier picture. Correlations in conductance values are observed, but they can be more or less strong and either positive or negative depending on ion channel subtypes and neuron subtypes \citep{tapia2018neurotransmitter, iacobas2019coordinated, kodama2020graded, khorkova2007neuromodulators}. In addition, correlations in both ion channel expression and conductance values can be activity- and neuromodulation-dependent \citep{santin2019membrane, temporal2012neuromodulation}. The emergence of negative correlations in conductance values poses the question of what potentially complex mechanism might link channel expression and conductance value. In this work, we tackle this question by analyzing how positive and negative conductance correlations arise in highly degenerate parameter sets of two different conductance-based models. We show that pairwise correlations in channel conductance are the result of two interfering mechanisms. Such interference is activity-dependent, which results in activity-dependent correlation levels. 
Second, our understanding of how ion channels shape neuronal activity remains largely qualitative to date. The lack of concrete mechanistic understanding makes it extremely difficult to quantify how specific changes in ion channel densities would affect neuronal output, which in turn makes the study of reliable neuromodulation an arduous task. Here, we provide such a mechanistic understanding through a dimensionality reduction analysis of the two degenerate parameter sets. The geometry of the principal components found by dimensionality reduction methods is fully explained by the geometry of the sensitive directions in the maximal conductance space, as revealed by using feedback control ideas \citep{drion2015dynamic}. This analysis permits to derive a simple, physiologically plausible rule for reliable neuromodulation in highly degenerate neurons. 

\clearpage
\section*{\textbf{RESULTS}}
\subsection*{\textbf{Neuronal degeneracy in conductance-based models is associated with variable pairwise correlations in channel conductances}}

\begin{figure}[ht!]
\centering
\includegraphics[width=\linewidth]{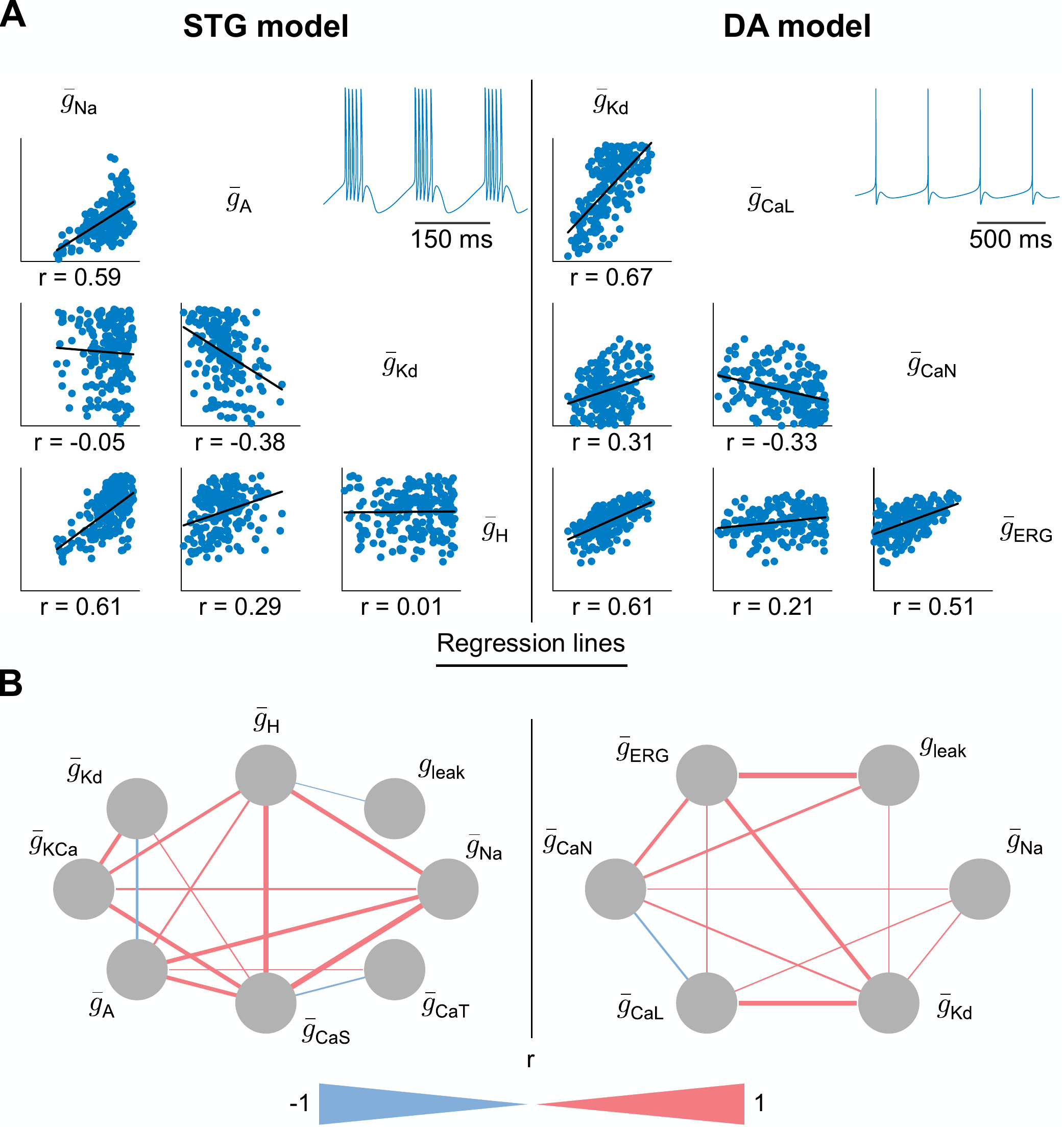}
\caption{\textbf{Neuronal degeneracy in conductance-based models is associated with variable pairwise correlations in channel conductances.}\\
(A) Scatter plot matrices of random sampling populations in the conductance spaces for STG model (left) and DA model (right), along with regression lines. The pairs depicted here do not represent all conductances of the models and are chosen randomly to illustrate the variable correlations, expressed by the Pearson correlation coefficient (r). All conductances are expressed in \si{mS\per cm^2}. Each bottom-left corner of every scatter plot represents the origin of the conductance space, and ranges can be found in STAR\textbigstar METHODS.\\
(B) Correlation graphs of all conductances of the random sampling populations for the STG model (left) and DA model (right). A blue (red) line indicates a negative (positive) pairwise correlation. The thickness of the line represents the absolute value of the correlation. Correlations below a certain threshold, corresponding to the inverse of the number of conductances in the considered model, are not shown.}
\label{fig:fig1}
\end{figure}

We first created variable sets of conductances leading to stable firing patterns in two different neuron conductance-based models (Figure \ref{fig:fig1}): a stomatogastric (STG) neuron model \citep{liu1998model} (left) and a dopaminergic (DA) neuron model (adapted from \cite{qian2014mathematical}) (right). All simulations and analyses were performed on these two different models to avoid uncovering model-specific features, but rather focus on general properties. Each parameter set was created through a random sampling followed by a post-processing procedure that selected models sharing specific firing pattern characteristics \citep{prinz2004similar}. Each model was first studied in its nominal firing pattern: burst firing for the STG neuron model, and slow tonic spiking for the DA neuron model (see STAR\textbigstar METHODS). An example of each firing pattern is shown at the top, right of each panel of Figure \ref{fig:fig1}A. 

Figure \ref{fig:fig1}A shows a scatter plot matrix of ion channel maximal conductances for a subset of ion channel types in both models, as well as the correlation computed for each pair. In agreement with what has been observed in previous experimental and computational work \citep{goaillard2021ion, khorkova2007neuromodulators}, correlations can be highly variable between different pairs of conductances, some being strongly positively correlated (such as $\bar{g}_{\mathrm{Na}}$ and $\bar{g}_{\mathrm{A}}$ in STG model), some negatively correlated (such as $\bar{g}_{\mathrm{A}}$ and $\bar{g}_{\mathrm{Kd}}$ in STG model), and others seemingly uncorrelated. This highlights the strong degeneracy of both conductance-based models, although they both maintain their specific firing activity using different types of ion channels. 

To gain further insights into how conductances correlate to maintain a robust firing activity, we represent the pairwise correlations between all conductances using correlation graphs (Figure \ref{fig:fig1}B). Each node represents a conductance, the thickness of the edges connecting each node represents the correlation strength, and the color of each edge represents the correlation sign (red for positive and blue for negative). These two graphs show a similar trend for both models: correlations between ion channels are dominantly positive, but negative correlations also emerge in a small subset of conductance pairs. This observation is intriguing for two reasons. 

First, in order to maintain a similar firing activity, one would expect conductances that are sources of currents of the same sign to correlate negatively, whereas conductances that are sources of currents of the opposite sign to correlate positively. This would allow to maintain the global transmembrane current, hence excitability, at a steady level. This is not what is observed in Figure \ref{fig:fig1}B. If we take the example of $\bar{g}_{\mathrm{CaS}}$ in STG model, which is a source of inward current, it can either correlate  negatively or positively with other sources of inward currents ($\bar{g}_{\mathrm{CaT}}$ and $\bar{g}_{\mathrm{Na}}$, respectively). Likewise, outward current sources can correlate both negatively or positively with other outward sources (i.e. $\bar{g}_{\mathrm{Kd}}$ with $\bar{g}_{\mathrm{A}}$ and $\bar{g}_{\mathrm{KCa}}$ in STG model). The same observation can be made for the DA neuron model. 

Second, experimental studies on the correlation between ion channel mRNA and computational models of neuronal homeostasis have uncovered the existence and emergence of neuron-dependent, strictly positive correlations in channel densities \citep{goaillard2021ion, o2013correlations, tobin2009correlations}. A similar trend emerges from our data set, where the vast majority of correlations are indeed positive. But, in opposition to homeostasis models and in agreement with experimental data \citep{khorkova2007neuromodulators}, negative correlations are also observed, which suggests that correlations emerging from homeostatic rules are important to maintain a robust firing activity, but that some other mechanisms must be at play. 

\clearpage

\clearpage

\subsection*{\textbf{A few principal components capture neuronal degeneracy but do not single out channel functions}}

\begin{figure}[ht!]
\centering
\includegraphics[width=\linewidth]{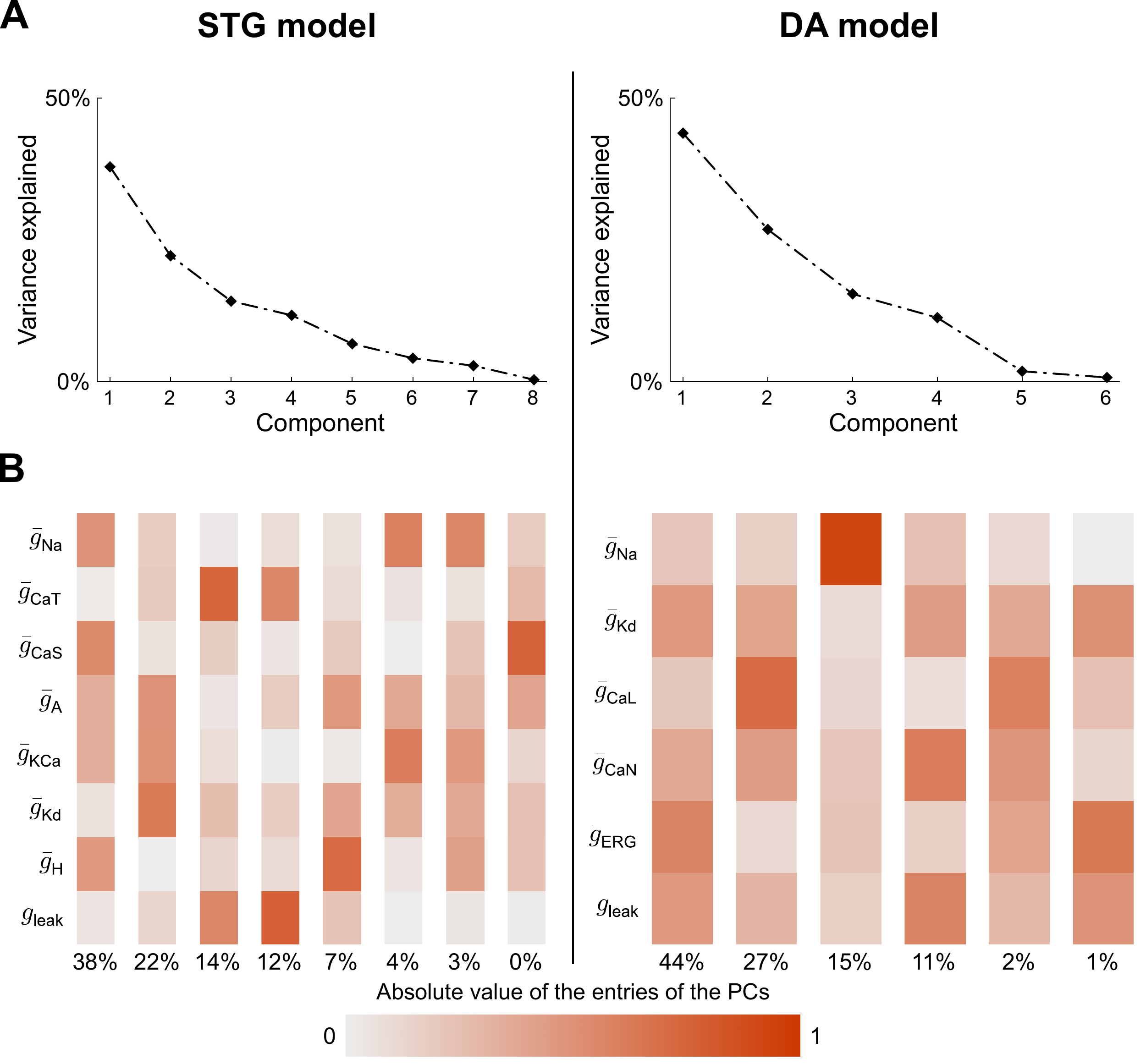}
\caption{\textbf{A few principal components capture neuronal degeneracy but do not single out channel functions.}\\
(A) Scree plot of PCA applied to the conductance spaces of random sampling populations for STG model (left) and DA model (right).\\
(B) Absolute values of the entries of the PCs in the conductance space for the STG model (left) and the DA model (right).}
\label{fig:fig2}
\end{figure}

As pairwise correlations between conductances alone did not provide much insight into how ion channels correlate to maintain a robust firing activity, we performed a principal component analysis (PCA) of both random sampling sets in an attempt to uncover low-dimensional subspaces in the data. We observed that a limited number of principal components, 4 for the STG model and 3 for the DA model, accounts for more than 80\% of the total variances in the data (Figure \ref{fig:fig2}A). We chose to focus our analysis on these significant principal components. The first principal component accounted for around 40\% of variance in both models. This observation is encouraging, as it shows that the mechanisms that drive conductance joint distribution in neuron models are low-dimensional, which is key for interpretability. 

We secondly extracted the contribution of each conductance in each of the principal components, with the hope to observe a pattern that would allow us to make predictions on the biophysics behind these components (Figure~\ref{fig:fig2}B). The results were however difficult to interpret, as a variety of conductances contributed to the different principal components for both models. Moreover, conductances that contributed much to the first principal component in one model did not in the other (see for instance the role of $\bar{g}_{\mathrm{Na}}$ or $g_{\mathrm{leak}}$ in both models), which did not permit to extract a model-independent rule. Although this last observation might seem unsurprising, as both models relate to different neurons exhibiting different firing patterns from different ion channels, we still aim to find some common, general mechanisms that might rule degeneracy in ion channel conductances.

\clearpage

\clearpage

\subsection*{\textbf{The dominant principal component captures homogeneous scaling of maximal conductances}}

\begin{figure}[ht!]
\centering
\includegraphics[width=\linewidth]{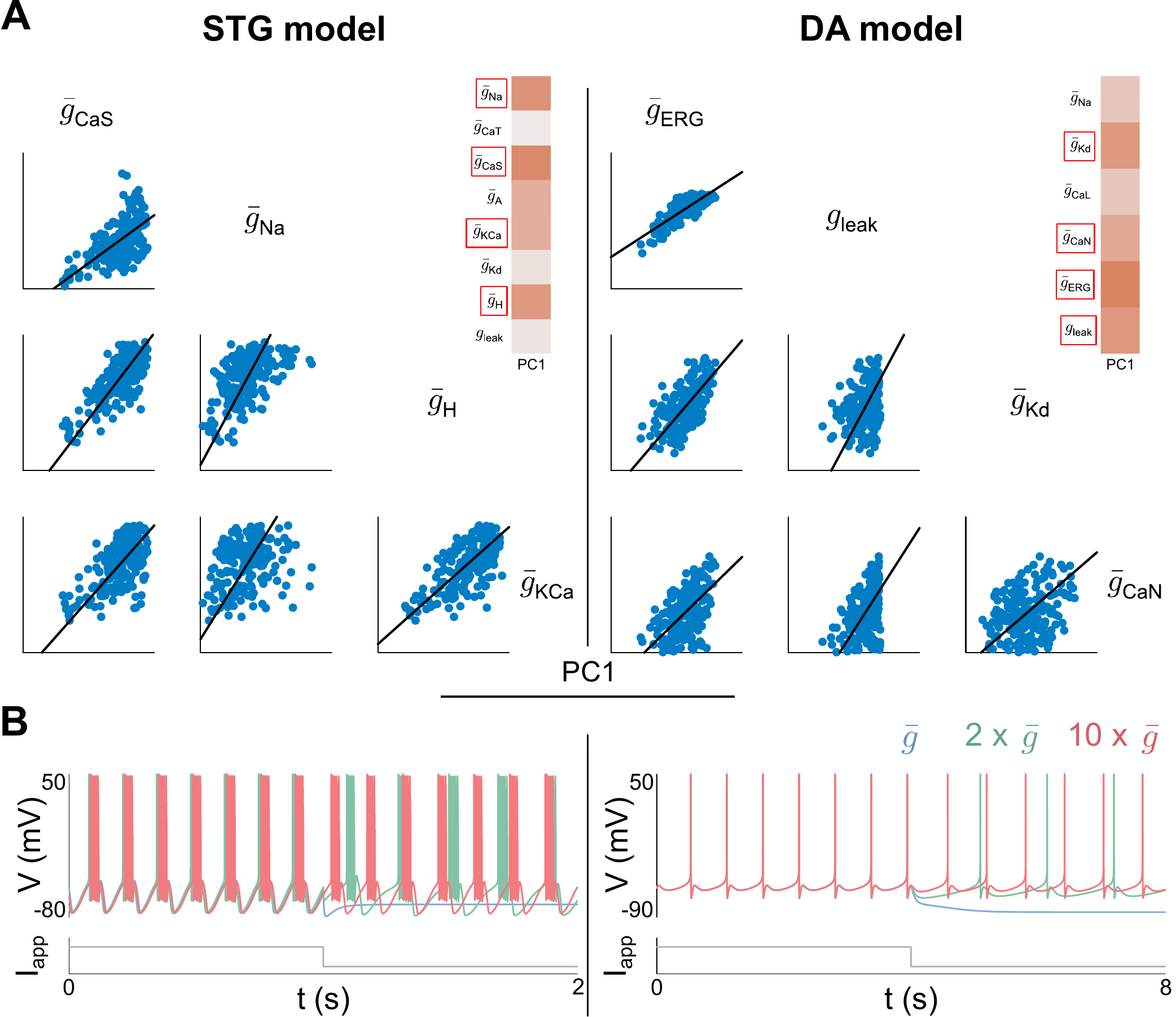}
\caption{\textbf{The dominant principal component captures homogeneous scaling of maximal conductances.}\\
(A) Scatter plot matrices of random sampling populations in the conductance spaces for STG model (left) and DA model (right) along with the direction of PC1. The scatter plots shown are associated with the conductances having largest entries (in absolute value) in the first PC.
All conductances are expressed in \si{mS\per cm^2}. Each bottom-left corner of every 2D subspace represents the origin of the conductance space, with ranges detailed in STAR\textbigstar METHODS.\\
(B) Simulations illustrating the effect of homogeneous scaling for the STG model (left) and the DA model (right). A random model from the scatter plot in (A) receives an inhibitory input (blue). The same experiment is then conducted with all conductances multiplied by 2 (green) and 10 (red).}
\label{fig:fig3}
\end{figure}

As the first principal component (PC1) accounted for a large portion of the variability in the data for both models (around 40 \%), we further analyzed its role by plotting the scatter plots of conductance values for a subset of four conductances that play a dominant role in PC1 (Figure \ref{fig:fig3}A). Interestingly, these scatter plots show that all conductances that play a significant role in PC1 show a strong positive correlation with each other in both models. This picture is very reminiscent of what is observed in channel mRNA data or the resulting channel correlations emerging from models of neuronal homeostasis \citep{o2013correlations, o2014cell, marder2006variability}. In particular, such positive correlations follow a direction passing roughly through the origin. 

Such direction is close to the homogeneous scaling direction in the maximal conductances. The direction of homogeneous scaling corresponds to the total least squares regression direction without intercept, \textit{i.e.}, to the direction connecting the origin of the conductance space to the center of mass of the degeneracy set. This center of mass represents the means of every type of conductance across the population.
While pairwise homogeneous scaling is only evident in a subset of ion channels, this observation extends to the entire conductance space. The alignment between PC1 and homogeneous scaling in the full conductance space was robustly confirmed in both the STG and DA models, with a notable 0.8 alignment in the former and a remarkable 0.9 alignment in the latter. This alignment was computed as the cosine of the angle between PC1 and homogeneous scaling direction.
Alternatively, it can be interpreted as the cosine of the angle formed by these two directions in the high dimensional space of conductances.

The dominant role of homogeneous scaling of conductances in neuronal degeneracy can be understood by its functional significance. Such homogeneous scaling can emerge from homeostatic models of ion channel expression, where the slope between a pair of conductances correlates with neuronal activity type \citep{o2014cell}. This slope is determined by the ratio of regulation time constants. Homogeneous scaling also permits to modulate the neuron response to external inputs while maintaining its intrinsic firing pattern unaffected. Indeed, increasing all conductances by a common factor permits to increase the global membrane permeability, hence decreasing its responsiveness to external input through a decrease in its input resistance (Figure \ref{fig:fig3}B). At the same time, it does not affect the ratio between channel conductances, thus maintaining firing activity. Homogeneous scaling therefore has a critical role in excitability modulation and homeostasis. 

\clearpage

\clearpage
\subsection*{\textbf{An alternative approach to build degenerate parameter sets permits to separate the effect of homogeneous scaling from other sources of degeneracy.}}




Analysis of the next meaningful principal components (PC2, PC3, and PC4 in STG model, and PC2 and PC3 in DA model) should permit to understand the physiological origin of most of the remaining variance in the data. However, these principal components have highly variable slopes in the different conductance planes, which makes the analysis less straightforward than for PC1. The effect of homogeneous scaling is intertwined with the other potential origins of degeneracy in the neuron model populations, which blur the picture even more.

To circumvent this problem, we constructed a new dataset that allowed us to separate the effect of homogeneous scaling from other potential effects. This dataset was constructed by leveraging the concept of dynamic input conductances (DIC) \citep{drion2015dynamic}, which provides a way to link channel conductance ratios with firing activity. In short, it was shown that the dynamical effects of ion channel gating on neuron activity could be captured by a few voltage-dependent conductances (DIC) acting on separate timescales. For a bursting neuron, three timescales are sufficient: a fast timescale characterizing spike upstroke, a slow timescale characterizing spike downstroke, neuron excitability type and rest-spike bistability, and an ultraslow timescales characterizing burst parameters such as period and duty cycle. The value of each DIC at threshold potential on each timescale determines firing activity, and each parameter sets leading to similar DIC values leads to similar firing activities. We exploited this last property to build degenerate parameter sets by identifying directions of zero sensitivity in the maximal conductance space, \textit{i.e.}, directions along which changes in maximal conductances do not affect DIC values at threshold, hence lead to similar spiking behavior \citep{drion2015dynamic}.
In practice, we created datasets of similar firing patterns by allowing for randomness in all conductances but one per timescale, and adapting the remaining conductances to ensure that DIC values are maintained constant (see STAR\textbigstar METHODS for further details). Importantly, in order to be able to separate the effect of homogeneous scaling from other sources of ion channel degeneracy, we normalized DIC values by $g_{\mathrm{leak}}$. This normalization permits to create variable conductance ratios that barely affects homogeneous scaling, which is itself mostly captured through variability in $g_{\mathrm{leak}}$, the leak conductance being the dominant current source below threshold potential.

\begin{figure}[ht!]
\centering
\includegraphics[width=\linewidth]{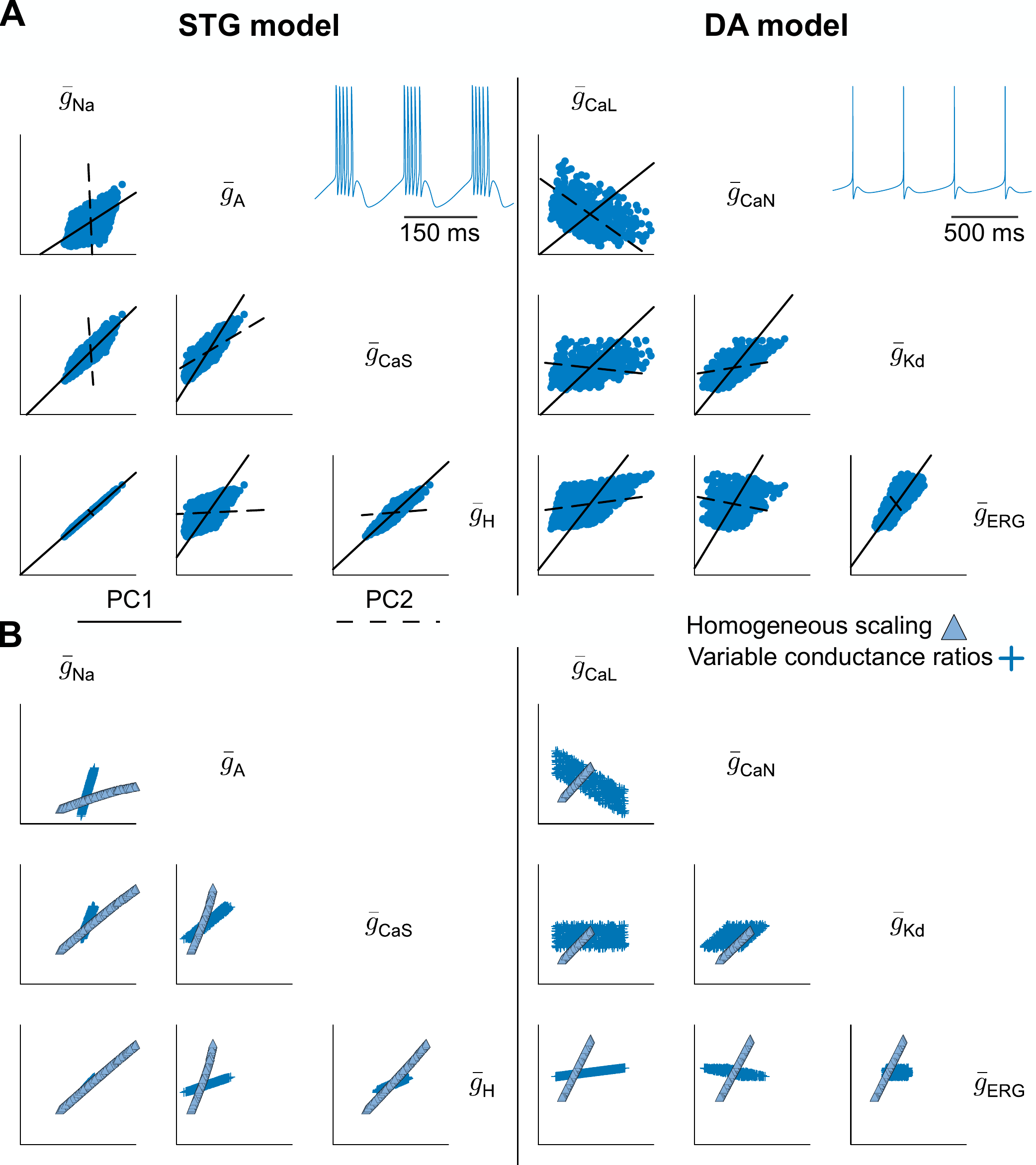}
\caption{\textbf{An alternative approach to build degenerate parameter sets permits to separate the effect of homogeneous scaling from other sources of degeneracy.}\\
(A) Scatter matrices of custom generated populations in the conductance spaces for STG model (left) and DA model (right) along with the directions of PC1 and PC2. The 2D subspaces shown here do not represent all conductances of the models and are randomly chosen. All conductances are expressed in \si{mS\per cm^2}. Each bottom-left corner of every 2D subspace represents the origin of the conductance space, and ranges can be found in STAR\textbigstar METHODS.\\
(B) Scatter matrices of custom generated populations in the conductance spaces for STG model (left) and DA model (right), isolating the effects of homogeneous scaling only (resp. variability in conductance ratios) shown as triangles (resp. crosses). The 2D subspaces are the same as in (A). All conductances are expressed in \si{mS\per cm^2}. Each bottom-left corner of every 2D subspace represents the origin of the conductance space, and ranges can be found in STAR\textbigstar METHODS.}
\label{fig:fig4}
\end{figure}

The dataset constructed using this approach created neurons exhibiting similar firing activities (see Supplementary Figure in STAR\textbigstar METHODS) and showed close qualitative similarities with the dataset produced through random sampling in both models: pairwise correlations in channel expression are highly variable between channel pairs, with a dominance of positive correlations but also the existence of negative correlations, the first principal component align with homogeneous scaling, and the second principal component has highly variable slopes in the different conductance planes (Figure \ref{fig:fig4}A). 

To elucidate the source of the second principal component, the dataset was partitioned into two subsets (Figure \ref{fig:fig4}B): one characterized by variability solely in $g_{\mathrm{leak}}$ (triangles in Figure \ref{fig:fig4}B) and another exhibiting variability exclusively in voltage-gated conductance ratios along DIC zero sensitivity directions (crosses in Figure \ref{fig:fig4}B). Variability in $g_{\mathrm{leak}}$ only creates a degenerate dataset with strong, strictly positive correlations between conductance pairs, which isolates the effect of homogeneous scaling in channel conductances. Regression slopes of these subsets strongly align with the first principal component of the full dataset. 

Variability limited to voltage-gated conductances (and fixed $g_{\mathrm{leak}}$) creates a degenerate dataset that also shows strong pairwise correlations. However, these correlations can be either positive or negative, and their regression slopes do not intersect the origin. Within this subset, the correlation between pairs of conductances arises from their distinct roles in shaping DIC values at threshold, and the slow DIC in particular. Indeed, we found that the effect of the slow DIC was dominant in our dataset, as the slow DIC rules spiking to bursting transitions through changes in neuron excitability type and the regulation of rest-spike bistability (see STAR\textbigstar METHODS for further details). Channels that have an opposite effect on the slow DIC show a positive correlation ($\bar{g}_{\mathrm{CaS}}$ and $\bar{g}_{\mathrm{A}}$ in STG for instance), whereas channels that have a similar effect show a negative correlation ($\bar{g}_{\mathrm{CaL}}$ and $\bar{g}_{\mathrm{CaN}}$ in DA model for instance). The regression slopes within this subset strongly align with the second principal component (PC2) of the complete dataset (compare PC2 in Figure \ref{fig:fig4}A with crosses in Figure \ref{fig:fig4}B). 

This alternative approach to build degenerate parameter sets shows that variable pairwise correlations in channel conductances could result from the interaction of two distinct mechanisms: homogeneous scaling, which maintain the ratio between ion channel conductances, and degenerate conductance ratio that lead to similar DIC values, hence similar membrane dynamical properties. 

\clearpage

\subsection*{\textbf{Secondary principal components also capture degenerate conductance ratios that maintain DIC values in the original random sampling dataset}}

\begin{figure}[ht!]
\centering
\includegraphics[width=\linewidth]{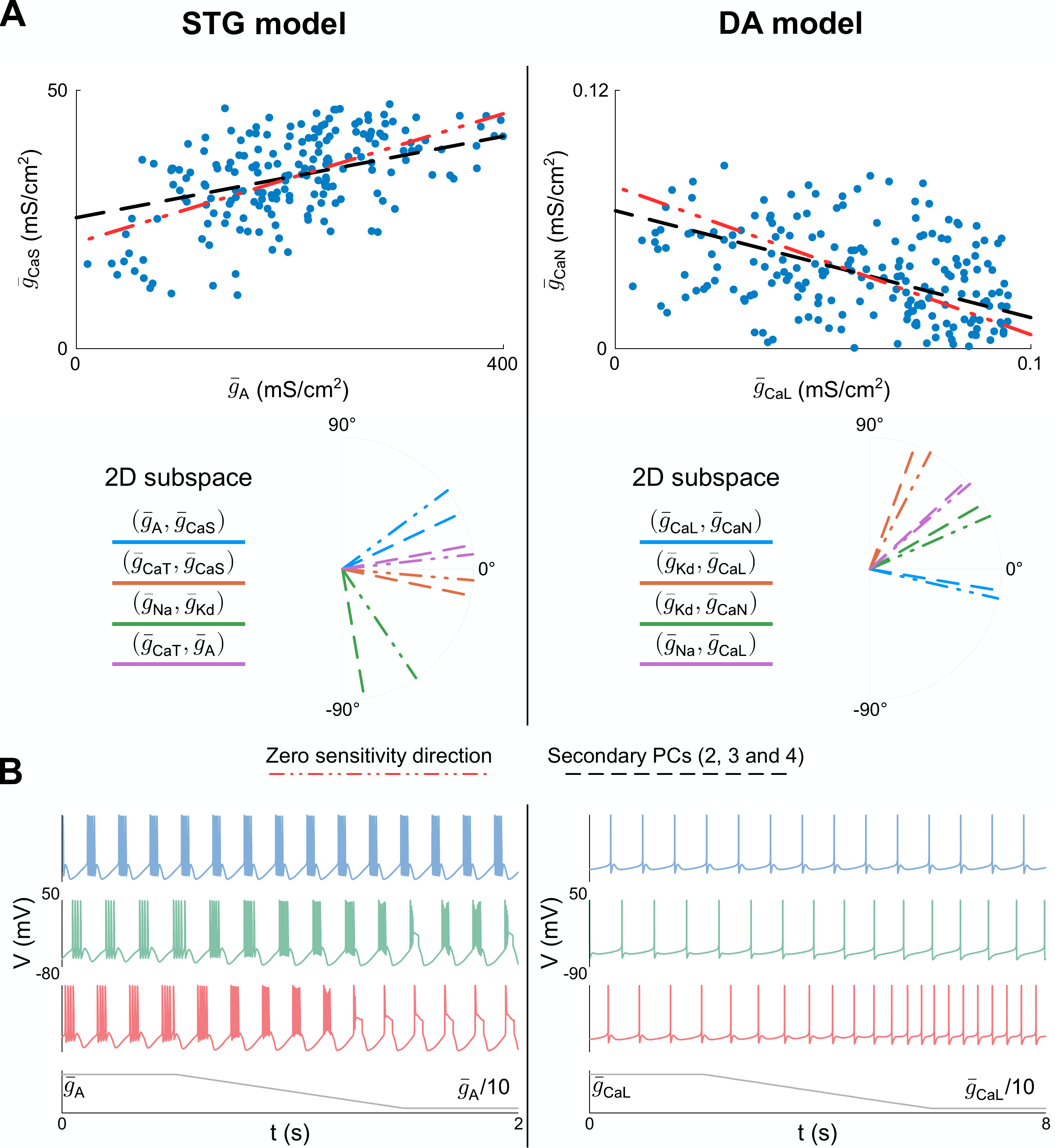}
\caption{\textbf{Secondary principal components also capture degenerate conductance ratios that maintain DIC values in the original random sampling dataset.}\\
(A) Scatter plots (top) of random sampling populations in the $(\bar{g}_\mathrm{A},\bar{g}_\mathrm{CaL})$ 2D subspace for STG model (left) and the $(\bar{g}_\mathrm{CaL},\bar{g}_\mathrm{CaN})$ 2D subspace for DA model (right), along with polar plots of PC2 (dashed line) and the zero sensitivity direction (dash-dotted line) in randomly chosen 2D subspaces of the conductance space (bottom).\\
(B) Cartoon simulations illustrating the effect of conductance ratios for the STG model (left) and the DA model (right). Three random points (blue, green, and red) in (A) undergo a disturbance in one conductance by dividing $\bar{g}_\mathrm{A}$ (STG model) and $\bar{g}_\mathrm{CaL}$ (DA model) by 10.
}
\label{fig:fig5}
\end{figure}

We then verified if variability in conductance ratio leading to similar DIC values was also a dominant source of degeneracy in the original random sampling dataset by computing zero sensitivity directions of slow DIC in both STG and DA neuron models and compare these directions with the secondary principal components (PC2, PC3, and PC4). In both models, the zero-sensitivity directions strongly align with one of the secondary principal components in the random sampling set (Figure \ref{fig:fig5}A). This confirmed that the second origin of degeneracy in ion channel expression can be explained by the existence of degenerate conductance ratios that create similar membrane dynamical properties.

Degeneracy in conductance ratios is also of functional significance for robust neuronal signaling. Relying on different conductance ratios to create a similar firing activity permits to create heterogeneity in response to external perturbations such as changes in temperature or pH \citep{haley2018two, rinberg2013effects}, as well as specific ion channel blockades or dysfunction, which increases neuronal robustness (Figure \ref{fig:fig5}B). It also creates variable responses to exogenous neuroactive drugs, and allows for compensation during long-lasting drug exposure or genetic defect in specific channel expression.

\clearpage

\clearpage

\subsection*{\textbf{Variability from both homogeneous scaling and degenerate conductance ratios blurs the connection between conductance correlation and their function}}

\begin{figure}[ht!]
\centering
\includegraphics[width=\linewidth]{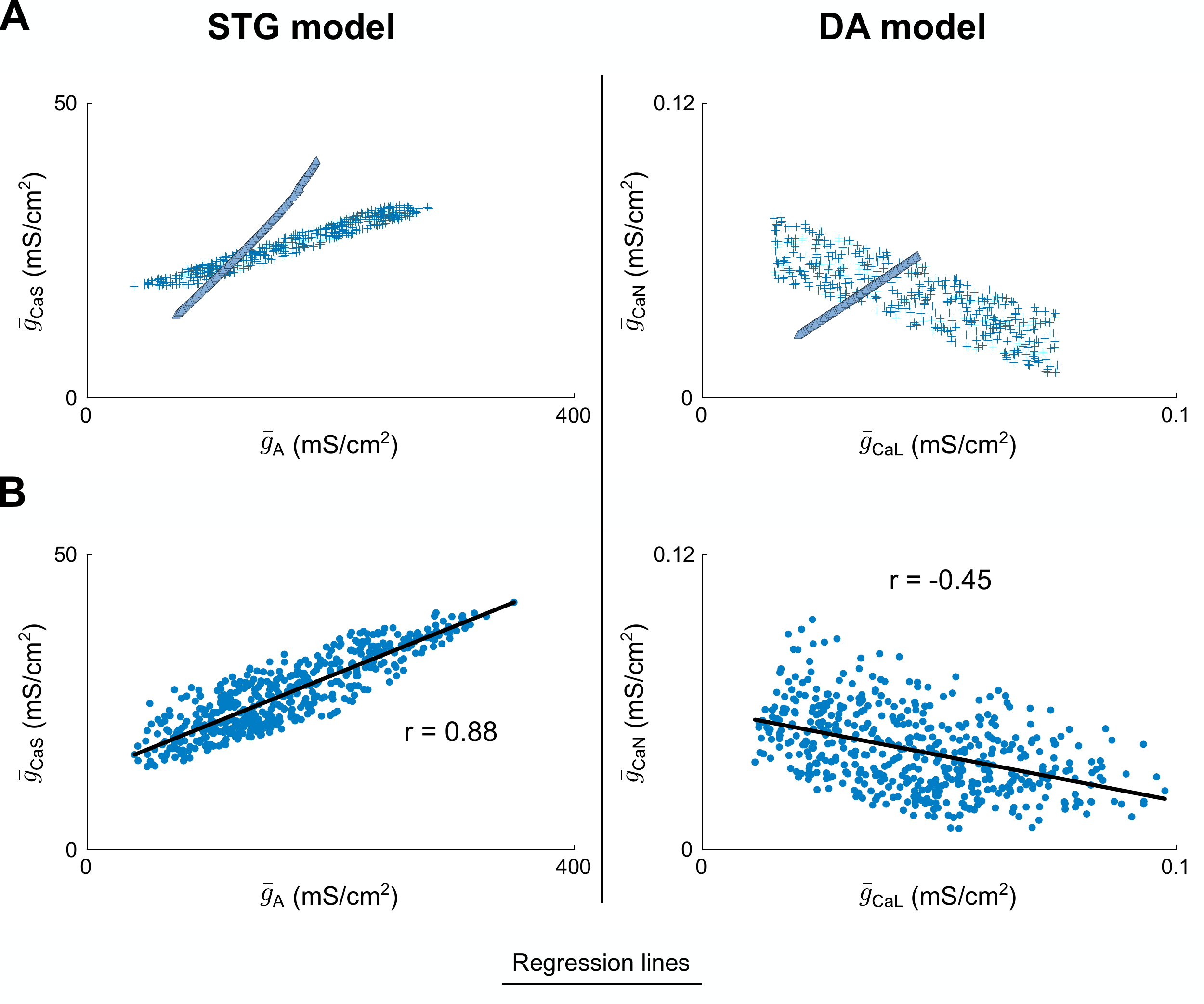}
\caption{\textbf{Variability from both homogeneous scaling and degenerate conductance ratios blurs the connection between conductance correlation and their function.}\\
(A) Scatter plots of custom generated populations separated into triangles (homogeneous scaling only) and crosses (variability in conductance ratios only) in the $(\bar{g}_\mathrm{A},\bar{g}_\mathrm{CaL})$ 2D subspace for STG model (left) and the $(\bar{g}_\mathrm{CaL},\bar{g}_\mathrm{CaN})$ 2D subspace for DA model (right). \\
(B) Scatter plots of the full variability custom generated populations in the same 2D subspace as in (A) for both the STG model (left) and the DA model (right), along with regression lines and Pearson correlation coefficients.}
\label{fig:fig6}
\end{figure}

Our analysis so far showed that variability from homogeneous scaling creates strong positive correlations in channel conductances. On the other hand, variability in voltage-gated conductance ratios also leads to strong correlations in channel conductances, but which can either be positive or negative depending on the channel pairs (Figure \ref{fig:fig6}A). When both variability types are present within a neuron population, these two correlation mechanisms interfere with each other to create highly variable correlation levels between channel pairs (Figure \ref{fig:fig6}B). If both variability types create positive correlations, the interference is minimal, and the global correlation in channel conductance remains strong (Figure \ref{fig:fig6}, left). If variability in conductance ratios creates a negative correlation, the interference is consequential, and the global correlation in channel conductance becomes weak (Figure \ref{fig:fig6}, right). This observation is of interest, as it shows that the variable pairwise correlation observed in channel conductance values originate from potentially competing effects rather than from an actual uncorrelated role in our datasets.

\clearpage

\subsection*{\textbf{Variability in pairwise correlations in conductance values is neuromodulation-dependent}}

\begin{figure}[ht!]
\centering
\includegraphics[width=0.9\linewidth]{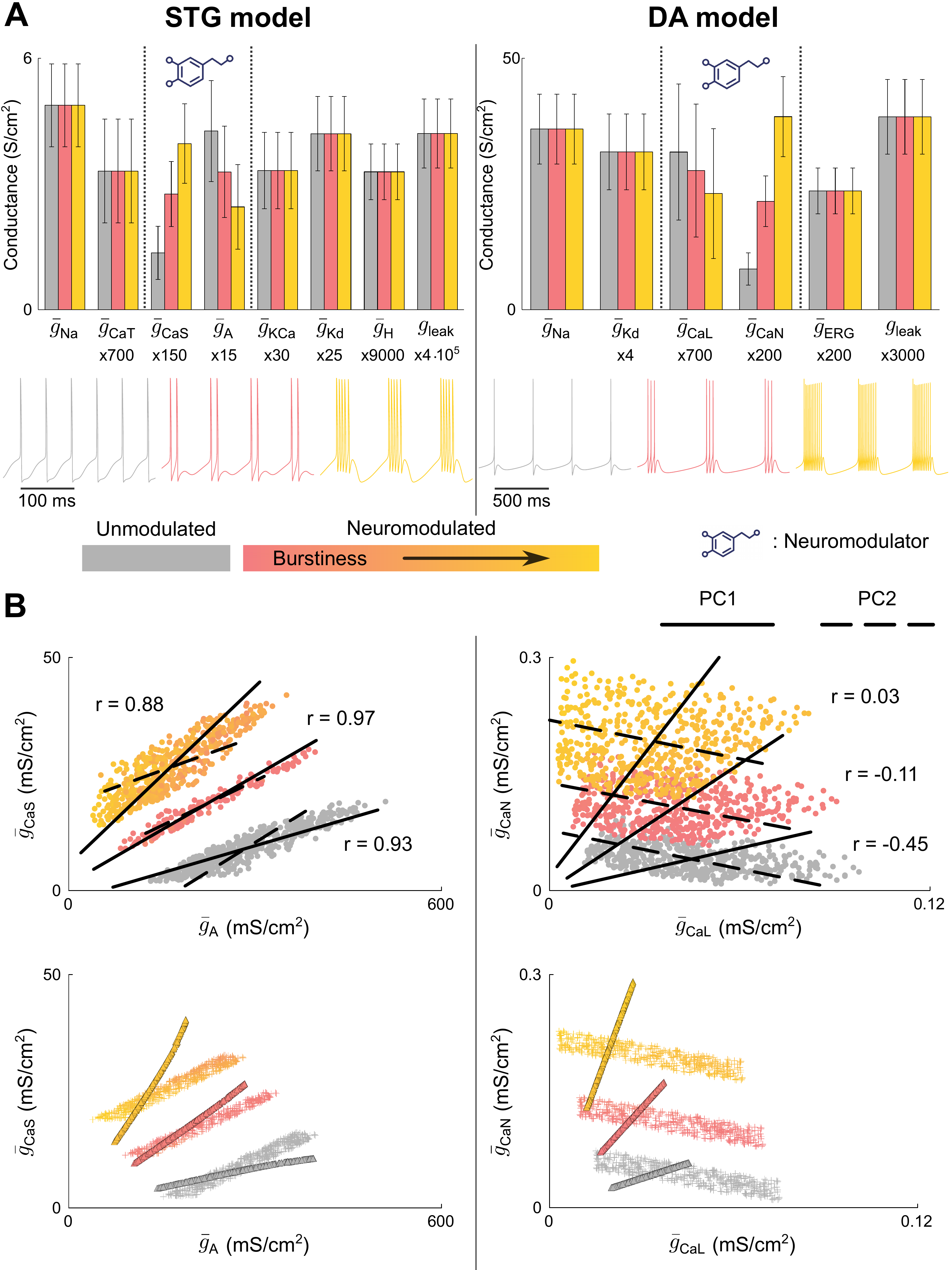}
\caption{\textbf{Variability in pairwise correlations in conductance values is neuromodulation-dependent.}\\
(A) Bar plot of conductance values for custom generated populations in the 3 phenotypes considered for STG model (right) and DA model (left).\\
(B) Scatter plots of full variability custom generated populations in the neuromodulated 2D space for both STG model (left) and DA model (right) across three neuromodulated states, along with PC1, PC2, and Pearson correlation coefficients.\\
(C) Scatter plots of separated custom generated populations in the neuromodulated 2D space for both STG model (left) and DA model (right) across three neuromodulated states.}
\label{fig:fig7}
\end{figure}

The variability in channel pairwise correlation level is therefore linked to the relative slope of the correlations created by both variability types, homogeneous scaling and degenerate conductance ratios. This has an interesting consequence when one studies the effect of neuromodulation on correlation in channel conductance. To illustrate this consequence, we performed a simple computational experiment where we neuromodulated the excitability state of both models from spiking to light bursting to strong bursting (Figure \ref{fig:fig7}). In both cases, the neuromodulator affects the maximal conductance of two channel types: $\bar{g}_{\mathrm{A}}$ and $\bar{g}_{\mathrm{CaS}}$ in STG model, and $\bar{g}_{\mathrm{CaL}}$ and  $\bar{g}_{\mathrm{CaN}}$ in DA model (Figure \ref{fig:fig7}A).Those conductances are known to affect the burstiness of the respective neuron models. To create robust neuromodulation in degenerate neurons, we modulated the datasets of Figure \ref{fig:fig4}A by modifying the target threshold value for the slow DIC and used the algorithm of \cite{drion2015dynamic} to compute the neuromodulated conductance values for each neuron of the dataset (see STAR\textbigstar METHODS). The resulting data points are shown in the scatter plots at the top of Figure \ref{fig:fig7}B. The dot color quantifies neuron burstiness, showing that the three firing patterns are robustly attained and well separated. 

In both models, neuromodulation of neuron excitability strongly affects the level of pairwise correlations (Figure \ref{fig:fig7}B). In STG model, the correlation between $\bar{g}_{\mathrm{A}}$ and $\bar{g}_{\mathrm{CaS}}$ is strongly positive in spiking (r = 0.93), reaches a maximum in light bursting (r = 0.97), and decreases in strong bursting (r = 0.88). In DA model, the correlation between $\bar{g}_{\mathrm{CaL}}$ and $\bar{g}_{\mathrm{CaN}}$ is negative in spiking (r = -0.45), becomes less negative in light bursting (r = -0.11), and both conductances appear uncorrelated in strong bursting (r = 0.03). Pairwise correlations in ion channel conductances therefore appear neuromodulation-dependent.

The origin of these neuromodulation-dependent changes in pairwise correlations can be explained by plotting the first two principal components (PC1 and PC2) on the scatter plots and observing the effect of neuromodulation on these components. On the one hand, neuromodulation creates a rotation of PC1 around the origin, which affects its slope. In the projections of Figure \ref{fig:fig7}B, the slope of PC1 increases when neurons switch from spiking to bursting in both models. This effect is consistent with the results obtained from homeostatic models of ion channel expression \citep{o2014cell}. On the other hand, neuromodulation creates a translation of PC2, and the slope is barely affected. As a result, the relative slopes between PC1 and PC2 depend on neuron neuromodulation state, which affects the global correlation level.

In STG model, both PC components have a positive slope. In spiking, PC1 has a flatter slope than PC2, which slightly widens the data cloud. As the model switches to bursting mode, the slope of PC1 increases and both slopes become almost identical in light bursting. In this state, both principal components align, which creates a strong correlation between the channel pair. As the model further increases its burstiness, the slope of PC1 further increases and becomes greater than PC2. Both principal components disalign again and the correlation between the channel pair decreases. A similar observation can be drawn in DA model, except that here PC2 has a negative slope. As a result, PC1 and PC2 become more and more orthogonal as burstiness increases, which reduces correlation level, and even destroys channel pairwise correlation in a strong bursting state. 

As identified above, PC1 relates to the homogeneous scaling of conductances, whereas PC2 relates to the variability in the ratio between voltage-dependent conductances. To further demonstrate this link, we reproduced the three neuromodulation states in two subsets where we isolated variability coming from homogeneous scaling (triangles in the bottom panels of Figure \ref{fig:fig7}B) from variability in conductance ratios (crosses in the bottom panels of Figure \ref{fig:fig7}B). We used the same algorithm as for the full dataset to create robustly neuromodulated states. The results from both models clearly show that robust neuromodulation is achieved through a rotation of the data points in the conductance space if variability comes from homogeneous scaling, whereas it is achieved through a translation of the data points if variability is the ratio between voltage-dependent conductances. 

This observation can be interpreted physiologically and provides significant insights into the requirements for robust neuromodulation in variable neurons. If robust neuromodulation is achieved through a rotation in the conductance space, it means that the robust neuromodulation rule is multiplicative: $\bar{g}_{i,\mathrm{MOD}} = \alpha_i \cdot \bar{g}_{i,\mathrm{init}}$ where $\alpha_i$ is set by neuromodulator concentration. The rule is multiplicative in the case of variability through homogeneous scaling, because neurons having twice the maximal conductance values require twice the change in conductance to reach a similar firing pattern, due to the change in input resistance. If robust neuromodulation is achieved through a translation in the conductance space, it means that the robust neuromodulation rule is additive: $\bar{g}_{i,\mathrm{MOD}} = \bar{g}_{i,\mathrm{init}} + \beta_i$ where $\beta_i$ is as well set by neuromodulator concentration. The rule is additive in the case of variability in conductance ratios only, because a similar firing pattern is reached through a similar change in the normalized DIC values, which is created by the same change in maximal conductances. As a result, robust neuromodulation can be achieved through a simple, direct rule if only one type of variability is present in the neuronal population. A direct rule is however impossible to derive if both variability types are present in the population, which is likely considering the physiological significance of both types. Such a rule would indeed need to be both additive and multiplicative with a neuron-dependent ratio between both effects. Robust neuromodulation therefore requires an indirect rule involving a second messenger in highly degenerate neurons, which is precisely the mechanism observed in G protein-coupled receptor signaling. 

\clearpage

\subsection*{\textbf{A simple indirect rule for robust neuromodulation in highly degenerate neurons}}

\begin{figure}[ht!]
\centering
\includegraphics[width=\linewidth]{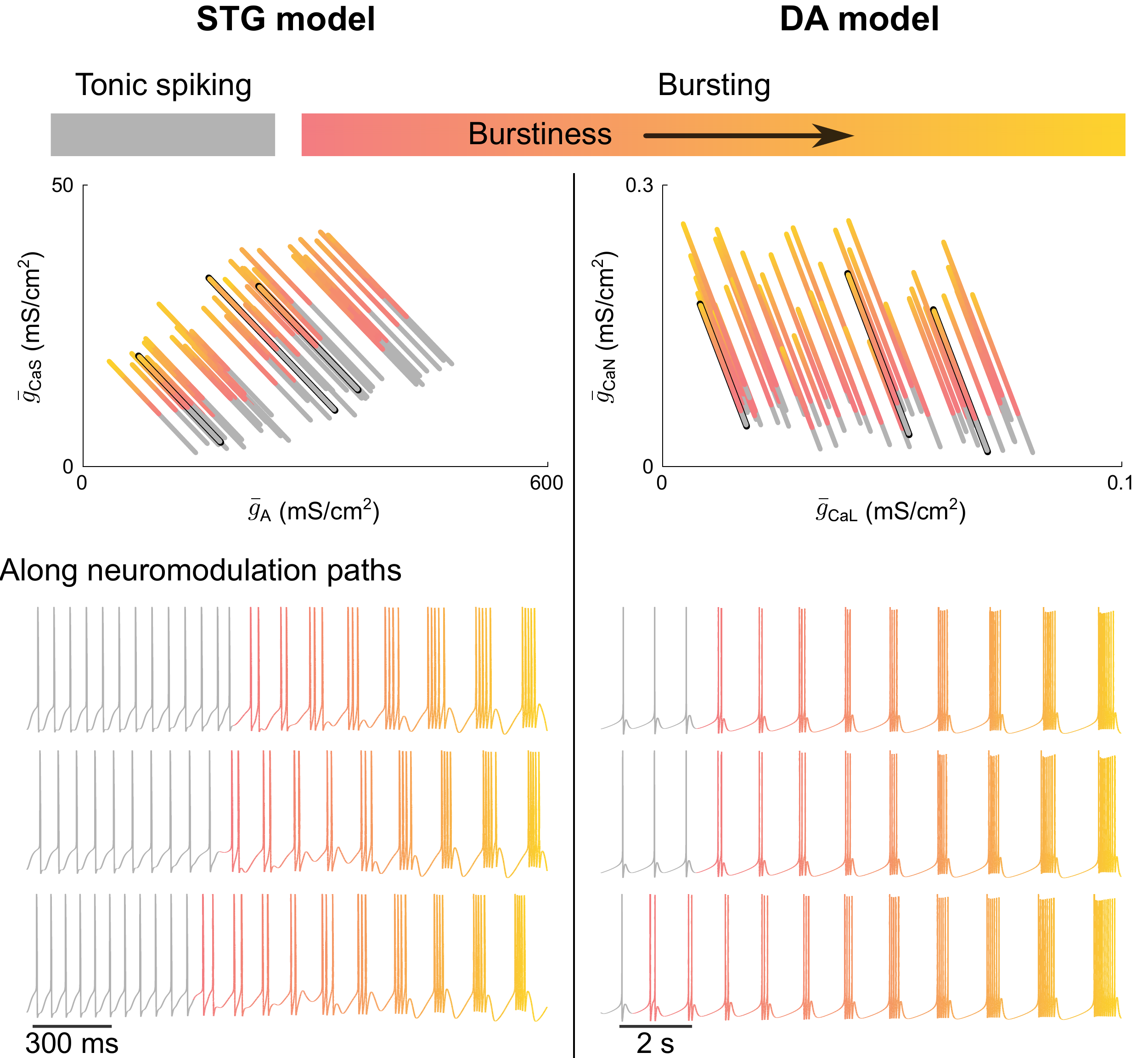}
\caption{\textbf{A simple indirect rule for robust neuromodulation in highly degenerate neurons.}\\
Neuromodulation paths of custom generated populations for a gradually continuous neuromodulation application in the neuromodulated 2D subspace (top). Each line corresponds to one neuron continuously undergoing neuromodulation. Three randomly chosen neuron voltage traces along their neuromodulation paths (bottom).}
\label{fig:fig8}
\end{figure}

We showed that robust neuromodulation in highly degenerate cells cannot rely on a simple rule directly targeting ion channels, but rather requires a more complex rule involving a second messenger. This raises the questions of how complex a rule for reliable neuromodulation should be, and whether a general, model-independent rule could be derived. To tackle this question, we used the algorithm developed above to construct reliable neuromodulation paths in degenerate neurons for both STG and DA models, moving from tonic spiking to bursting of increasing burstiness (Figure \ref{fig:fig8}). Similarly as above, the neuromodulation algorithm targeted $\bar{g}_{\mathrm{A}}$ and $\bar{g}_{\mathrm{CaS}}$ in STG model, and $\bar{g}_{\mathrm{CaL}}$ and $\bar{g}_{\mathrm{CaN}}$ in DA model. Many reliable neuromodulation paths could be achieved in both models using a simple rule whose objective is to increase the target threshold value for the slow DIC while moving from tonic spiking to bursting, while maintaining ultraslow DIC value constant to maintain spiking and bursting periods \citep{drion2015dynamic} (see STAR\textbigstar METHODS). Figure \ref{fig:fig8} plots the neuromodulation paths in the $(\bar{g}_{\mathrm{CaS}},\bar{g}_{\mathrm{A}})$ plane (resp. $(\bar{g}_{\mathrm{CaL}},\bar{g}_{\mathrm{CaN}})$ plane) for the STG model (resp. DA model) and examples of neuromodulated neuronal traces. 

Interestingly, although a simple direct rule cannot be used, the indirect rule resulted in linear neuromodulation paths for both models, where neuromodulation direction is constant and only varies between neurons of different types. The nonlinearity occurs in the distance the neuron has to move along that direction to switch activity, which is affected by parameter variability (see the variability in the color transitions of Figure \ref{fig:fig8} top). These results highlight the fact that, even in the case of maximal degeneracy in neuron parameters, the relative change in maximal conductances of ion channels targeted by the same neuromodulatory receptor can be hard-wired in a neuron type, creating a robust neuromodulation path. The role of the second messenger is then to control the movement along that neuromodulation path that would lead to the target activity, strongly reducing the complexity of the reliable neuromodulation process. Such control could for instance be implemented by sensing neuronal activity through intracellular calcium oscillations, as already suggested in homeostatic models \citep{liu1998model, o2014cell}, or by sensing membrane voltage \citep{santin2019membrane}, creating activity-dependent changes in targeted maximal conductances. Substantial evidence of such activity-dependent neuromodulation mechanisms involving intracellular calcium can be found in the experimental literature \citep{kramer1990activity, walters1984activity, marder2014neuromodulation, raymond1992learning}. 

\clearpage

\clearpage

\section*{\textbf{DISCUSSION}}


\subsection*{\textbf{Two physiologically relevant sources of neuronal variability rule ion channel degeneracy}}

Understanding how ion channel subsets shape neuronal excitability is critical to uncover how so many different neuron types emerge, as well as the mechanisms underlying reliable neuromodulation and variable neuronal response to neuroactive drugs \citep{amendola2012ca2, nadim2014neuromodulation, grashow2009reliable, schulz2006cellular, tobin2009correlations}. The connection between ion channels and neuronal signaling is however a complex task due to channel degeneracy, and despite considerable advances made on the subject through experimental, computational, and mathematical work, a  mechanistic understanding of ion channel variability and degeneracy in neurons remains elusive \citep{prinz2004similar, achard2006complex, alonso2019visualization, taylor2009multiple, swensen2005robustness}. In this work, we showed that neuronal variability can be separated into two quantifiable, physiological components: homogeneous scaling and variability in conductance ratios. 

Homogeneous scaling refers to the fact that neurons can exhibit a similar activity if the relative difference in their channel maximal conductances is similar for all channels expressed at the membrane, hence maintaining conductance ratios. Such property has been observed experimentally in channel expression and shown to emerge from homeostatic rules \citep{o2013correlations, o2014cell, marder2006variability}. In this case, intrinsic characteristics are maintained, but extrinsic excitability is altered due to differences in neuron input resistance. Variability in conductance ratios refers to the fact that neurons having a similar input resistance can exhibit a similar activity with different ratios in their voltage-gated conductances. In this case, intrinsic characteristics are maintained, but the response to perturbations such as temperature as well as channel blockade or dysfunction is altered due to differences in the relative role of each channel subtype on excitability.

Both sources of channel variability have physiological relevance. Homogeneous scaling is central for network homeostasis, as it permits to tune neuron input/output response while maintaining neuron intrinsic properties stable \citep{o2011neuronal}. Homogeneous scaling also permits to compensate for changes in membrane capacitance. Variability in conductance ratios on the other hand permits to improve robustness to external perturbations by creating an heterogeneous response to perturbations affecting specific channel functions at the network level \citep{drion2015ion}. It could also lead to variable inter-individual responses to neuroactive drugs. 

The contributions of variability from homogeneous scaling and conductance ratios are intertwined in any neuron degenerate dataset, making any quantification attempt difficult. Combining dimensionality reduction analysis and recent insights on reduced dynamics of conductance-based models, we were able to separate the contributions of both sources of variability, which permitted to construct a mechanistic understanding of how variable ion channels can lead to a specific neuronal activity. It permitted to understand the origin of ion channel variable pairwise correlations and derive a robust indirect rule for reliable neuromodulation in degenerate neurons.

\subsection*{\textbf{Variable channel correlations arise from the interference between homogeneous scaling and variability in conductance ratios}}

Separating the effects of homogeneous scaling and variability in conductance ratios permitted to analyze the role of both sources of variability on channel pairwise correlations. Homogeneous scaling creates strictly positive correlations between all ion channels, and different firing patterns/neuron subtypes lead to different regression slopes, as observed in channel expression data and homeostatic models of neuronal excitability \citep{o2014cell}. These positive correlations come from the passive role of ion channels on membrane properties through Ohm's law: increasing any channel conductance increases membrane permeability, which in turn decreases membrane input resistance. Other channels have therefore to increase their conductance to maintain their effect on membrane potential variations. 

Variability in conductance ratios on the other hand creates both positive and negative correlations between ion channel subsets, not all ion channels. Ion channels correlate to maintain neuronal dynamics if their gating, being either activation or inactivation, occurs on a similar timescale. The sign of the correlation is determined by the relative feedback provided by each channel gating on membrane potential variations, which is a key determinant of neuron dynamical properties as quantified by e.g. dynamic input conductances \citep{drion2015dynamic}. Specifically, activation of inward current and inactivation of outward current produce positive feedback, whereas activation of outward current and inactivation of inward current produce negative feedback. Two channels producing opposite feedbacks on a similar timescale will correlate positively (such as e.g. $\bar{g}_\mathrm{A}$ and $\bar{g}_\mathrm{CaS}$ in STG model), whereas two channels producing similar feedbacks will correlate negatively (such as e.g. $\bar{g}_\mathrm{CaL}$ and $\bar{g}_\mathrm{CaN}$ in DA model).

When both sources of variability exist in a neuron degenerate set, both types of correlations interfere with each other. When the correlation emerging from variability in conductance ratio is positive, both regression lines have a positive slope, creating an overall positive correlation whose correlation level depends on the alignment of the regression lines. However, when the correlation emerging from variability in conductance ratio is negative, both regression lines have opposite signs, which can lead to an uncorrelation between two conductances even though their role in neuron dynamics and passive properties strongly correlate. This situation could be indistinguishable from two channels that actually do not correlate due to a lack of action on a similar timescale. Variable correlations in channel conductances in a degenerate dataset therefore does not always relate to correlated or uncorrelated functions, but could also arise from highly correlated functions of opposite signs.

\subsection*{\textbf{The importance of indirect neuromodulation pathways for reliable neuromodulation in variable neurons.}}

One prominent question arising from channel degeneracy is how could neuromodulation be reliable across neurons when it acts on degenerate conductances \citep{nadim2014neuromodulation, grashow2009reliable, schulz2006cellular, marder2007understanding, marder2012neuromodulation, marder2014neuromodulation}. We showed that a simple direct rule for reliable neuromodulation could be derived if either homogeneous scaling or variability in conductance ratios existed in the dataset, but not both. Indeed, homogeneous scaling requires a simple multiplicative rule due to its effect on input resistance, whereas variability in conductance ratios requires an additive rule. A direct rule does not exist if both variability types are present in the population, as it would need to be both additive and multiplicative with a neuron-dependent ratio between both effects.

We showed that a simple indirect rule could produce reliable neuromodulation when both sources of variability are present in the dataset. This rule is indirect in the sense that it uses an intermediate signaling pathway to connect neuromodulation concentration with changes in channel conductances. In our computational study, this intermediate pathway encodes the values of the slow and ultraslow dynamic input conductances around threshold potential: neuromodulator concentration tunes the target values for both dynamic conductances, and a subset of ion channels are in turn modulated to reach these new targets. The presence of an intermediate messaging pathway is a core property of GPCR signaling, making such indirect rule physiologically plausible. Our work provides a quantitative framework that provides a new angle of attack to study how intermediate signaling pathways could lead to reliable neuromodulation in degenerate neurons. 

\section*{\textbf{STAR\textbigstar METHODS}}
Methods are briefly provided in the online version of this paper and include the following:
\begin{itemize}
    \item RESOURCE AVAILABILITY
    \begin{itemize}
        \item Lead contact
        \item Materials availability
        \item Data and code availability
    \end{itemize}
    \item CONDUCTANCE-BASED MODELS
    \item METHOD DETAILS
    \begin{itemize}
        \item Random sampling sets
        \item An efficient method to build degeneracy sets that allows to remove the effect of homogeneous scaling.
        \item Neuromodulation algorithm
    \end{itemize}
\end{itemize}

Mathematical details and experimental parameters can be found with this article online [DOI].

\section*{\textbf{ACKNOWLEDGMENTS}}
Arthur Fyon is a Research Fellow of the "Fonds de la Recherche Scientifique - FNRS". This work was supported by the Belgian Government through the FPS Policy and Support (BOSA) grant NEMODEI.

\section*{\textbf{AUTHOR CONTRIBUTIONS}}
A.Fy.\ and G.D.\ conceived and designed the research. 
A.Fy.\ and G.D.\ developed the model code. 
A.Fy.\ performed the simulations. 
A.Fy.\ analyzed the data. 
A.Fy., A.Fr., P.S., and G.D.\ interpreted the results of simulations. 
A.Fy.\ prepared figures. 
A.Fy.\ drafted the manuscript. 
A.Fy., A.Fr., P.S., and G.D.\ edited the manuscript. 
A.Fy., A.Fr., P.S., and G.D.\ approved the final version of the manuscript.

\section*{\textbf{DECLARATION OF INTERESTS}}
The authors declare no competing interests.

\section*{\textbf{DECLARATION OF GENERATIVE AI AND AI-ASSISTED TECHNOLOGIES IN THE WRITING PROCESS}}
During the preparation of this work, the authors used ChatGPT in order to improve language and readability of figure captions. After using this tool, the authors reviewed and edited the content as needed and take full responsibility for the content of the publication.

\clearpage
\section*{\textbf{REFERENCES}}
\bibliographystyle{apalike}
\bibliography{References}

\begin{thebibliography}{}

\bibitem[Achard and De~Schutter, 2006]{achard2006complex}
Achard, P. and De~Schutter, E. (2006).
\newblock Complex parameter landscape for a complex neuron model.
\newblock {\em PLoS computational biology}, 2(7):e94.

\bibitem[Alonso and Marder, 2019]{alonso2019visualization}
Alonso, L.~M. and Marder, E. (2019).
\newblock Visualization of currents in neural models with similar behavior and
  different conductance densities.
\newblock {\em Elife}, 8:e42722.

\bibitem[Amendola et~al., 2012]{amendola2012ca2}
Amendola, J., Woodhouse, A., Martin-Eauclaire, M.-F., and Goaillard, J.-M.
  (2012).
\newblock Ca2+/camp-sensitive covariation of ia and ih voltage dependences
  tunes rebound firing in dopaminergic neurons.
\newblock {\em Journal of Neuroscience}, 32(6):2166--2181.

\bibitem[Bezanson et~al., 2017]{bezanson2017julia}
Bezanson, J., Edelman, A., Karpinski, S., and Shah, V.~B. (2017).
\newblock Julia: A fresh approach to numerical computing.
\newblock {\em SIAM review}, 59(1):65--98.

\bibitem[Drion et~al., 2015a]{drion2015dynamic}
Drion, G., Franci, A., Dethier, J., and Sepulchre, R. (2015a).
\newblock Dynamic input conductances shape neuronal spiking.
\newblock {\em eneuro}, 2(1).

\bibitem[Drion et~al., 2015b]{drion2015ion}
Drion, G., O’Leary, T., and Marder, E. (2015b).
\newblock Ion channel degeneracy enables robust and tunable neuronal firing
  rates.
\newblock {\em Proceedings of the National Academy of Sciences},
  112(38):E5361--E5370.

\bibitem[Franci et~al., 2018]{franci2018robust}
Franci, A., Drion, G., and Sepulchre, R. (2018).
\newblock Robust and tunable bursting requires slow positive feedback.
\newblock {\em Journal of neurophysiology}, 119(3):1222--1234.

\bibitem[Goaillard and Marder, 2021]{goaillard2021ion}
Goaillard, J.-M. and Marder, E. (2021).
\newblock Ion channel degeneracy, variability, and covariation in neuron and
  circuit resilience.
\newblock {\em Annual review of neuroscience}, 44:335--357.

\bibitem[Grashow et~al., 2009]{grashow2009reliable}
Grashow, R., Brookings, T., and Marder, E. (2009).
\newblock Reliable neuromodulation from circuits with variable underlying
  structure.
\newblock {\em Proceedings of the National Academy of Sciences},
  106(28):11742--11746.

\bibitem[Haley et~al., 2018]{haley2018two}
Haley, J.~A., Hampton, D., and Marder, E. (2018).
\newblock Two central pattern generators from the crab, cancer borealis,
  respond robustly and differentially to extreme extracellular ph.
\newblock {\em Elife}, 7:e41877.

\bibitem[Iacobas et~al., 2019]{iacobas2019coordinated}
Iacobas, D.~A., Iacobas, S., Lee, P.~R., Cohen, J.~E., and Fields, R.~D.
  (2019).
\newblock Coordinated activity of transcriptional networks responding to the
  pattern of action potential firing in neurons.
\newblock {\em Genes}, 10(10):754.

\bibitem[Khorkova and Golowasch, 2007]{khorkova2007neuromodulators}
Khorkova, O. and Golowasch, J. (2007).
\newblock Neuromodulators, not activity, control coordinated expression of
  ionic currents.
\newblock {\em Journal of Neuroscience}, 27(32):8709--8718.

\bibitem[Kodama et~al., 2020]{kodama2020graded}
Kodama, T., Gittis, A.~H., Shin, M., Kelleher, K., Kolkman, K.~E., McElvain,
  L., Lam, M., and Du~Lac, S. (2020).
\newblock Graded coexpression of ion channel, neurofilament, and synaptic genes
  in fast-spiking vestibular nucleus neurons.
\newblock {\em Journal of Neuroscience}, 40(3):496--508.

\bibitem[Kramer and Levitan, 1990]{kramer1990activity}
Kramer, R.~H. and Levitan, I.~B. (1990).
\newblock Activity-dependent neuromodulation in aplysia neuron r15:
  intracellular calcium antagonizes neurotransmitter responses mediated by
  camp.
\newblock {\em Journal of neurophysiology}, 63(5):1075--1088.

\bibitem[Liss et~al., 2001]{liss2001tuning}
Liss, B., Franz, O., Sewing, S., Bruns, R., Neuhoff, H., and Roeper, J. (2001).
\newblock Tuning pacemaker frequency of individual dopaminergic neurons by kv4.
  3l and kchip3. 1 transcription.
\newblock {\em The EMBO journal}, 20(20):5715--5724.

\bibitem[Liu et~al., 1998]{liu1998model}
Liu, Z., Golowasch, J., Marder, E., and Abbott, L. (1998).
\newblock A model neuron with activity-dependent conductances regulated by
  multiple calcium sensors.
\newblock {\em Journal of Neuroscience}, 18(7):2309--2320.

\bibitem[Marder, 2012]{marder2012neuromodulation}
Marder, E. (2012).
\newblock Neuromodulation of neuronal circuits: back to the future.
\newblock {\em Neuron}, 76(1):1--11.

\bibitem[Marder and Bucher, 2007]{marder2007understanding}
Marder, E. and Bucher, D. (2007).
\newblock Understanding circuit dynamics using the stomatogastric nervous
  system of lobsters and crabs.
\newblock {\em Annu. Rev. Physiol.}, 69:291--316.

\bibitem[Marder and Goaillard, 2006]{marder2006variability}
Marder, E. and Goaillard, J.-M. (2006).
\newblock Variability, compensation and homeostasis in neuron and network
  function.
\newblock {\em Nature Reviews Neuroscience}, 7(7):563--574.

\bibitem[Marder et~al., 2014]{marder2014neuromodulation}
Marder, E., O'Leary, T., and Shruti, S. (2014).
\newblock Neuromodulation of circuits with variable parameters: single neurons
  and small circuits reveal principles of state-dependent and robust
  neuromodulation.
\newblock {\em Annual review of neuroscience}, 37:329--346.

\bibitem[Nadim and Bucher, 2014]{nadim2014neuromodulation}
Nadim, F. and Bucher, D. (2014).
\newblock Neuromodulation of neurons and synapses.
\newblock {\em Current opinion in neurobiology}, 29:48--56.

\bibitem[O'Leary et~al., 2013]{o2013correlations}
O'Leary, T., Williams, A.~H., Caplan, J.~S., and Marder, E. (2013).
\newblock Correlations in ion channel expression emerge from homeostatic tuning
  rules.
\newblock {\em Proceedings of the National Academy of Sciences},
  110(28):E2645--E2654.

\bibitem[O’Leary et~al., 2014]{o2014cell}
O’Leary, T., Williams, A.~H., Franci, A., and Marder, E. (2014).
\newblock Cell types, network homeostasis, and pathological compensation from a
  biologically plausible ion channel expression model.
\newblock {\em Neuron}, 82(4):809--821.

\bibitem[O’Leary and Wyllie, 2011]{o2011neuronal}
O’Leary, T. and Wyllie, D.~J. (2011).
\newblock Neuronal homeostasis: time for a change?
\newblock {\em The Journal of physiology}, 589(20):4811--4826.

\bibitem[Prinz et~al., 2004]{prinz2004similar}
Prinz, A.~A., Bucher, D., and Marder, E. (2004).
\newblock Similar network activity from disparate circuit parameters.
\newblock {\em Nature neuroscience}, 7(12):1345--1352.

\bibitem[Qian et~al., 2014]{qian2014mathematical}
Qian, K., Yu, N., Tucker, K.~R., Levitan, E.~S., and Canavier, C.~C. (2014).
\newblock Mathematical analysis of depolarization block mediated by slow
  inactivation of fast sodium channels in midbrain dopamine neurons.
\newblock {\em Journal of neurophysiology}, 112(11):2779--2790.

\bibitem[Raymond et~al., 1992]{raymond1992learning}
Raymond, J.~L., Baxter, D.~A., Buonomano, D.~V., and Byrne, J.~H. (1992).
\newblock A learning rule based on empirically-derived activity-dependent
  neuromodulation supports operant conditioning in a small network.
\newblock {\em Neural Networks}, 5(5):789--803.

\bibitem[Rinberg et~al., 2013]{rinberg2013effects}
Rinberg, A., Taylor, A.~L., and Marder, E. (2013).
\newblock The effects of temperature on the stability of a neuronal oscillator.
\newblock {\em PLoS computational biology}, 9(1):e1002857.

\bibitem[Santin and Schulz, 2019]{santin2019membrane}
Santin, J.~M. and Schulz, D.~J. (2019).
\newblock Membrane voltage is a direct feedback signal that influences
  correlated ion channel expression in neurons.
\newblock {\em Current Biology}, 29(10):1683--1688.

\bibitem[Schultz, 2007]{schultz2007multiple}
Schultz, W. (2007).
\newblock Multiple dopamine functions at different time courses.
\newblock {\em Annu. Rev. Neurosci.}, 30:259--288.

\bibitem[Schulz et~al., 2006a]{schulz2006cellular}
Schulz, D.~J., Baines, R.~A., Hempel, C.~M., Li, L., Liss, B., and Misonou, H.
  (2006a).
\newblock Cellular excitability and the regulation of functional neuronal
  identity: from gene expression to neuromodulation.
\newblock {\em Journal of Neuroscience}, 26(41):10362--10367.

\bibitem[Schulz et~al., 2006b]{schulz2006variable}
Schulz, D.~J., Goaillard, J.-M., and Marder, E. (2006b).
\newblock Variable channel expression in identified single and electrically
  coupled neurons in different animals.
\newblock {\em Nature neuroscience}, 9(3):356--362.

\bibitem[Schulz et~al., 2007]{schulz2007quantitative}
Schulz, D.~J., Goaillard, J.-M., and Marder, E.~E. (2007).
\newblock Quantitative expression profiling of identified neurons reveals
  cell-specific constraints on highly variable levels of gene expression.
\newblock {\em Proceedings of the National Academy of Sciences},
  104(32):13187--13191.

\bibitem[Swensen and Bean, 2005]{swensen2005robustness}
Swensen, A.~M. and Bean, B.~P. (2005).
\newblock Robustness of burst firing in dissociated purkinje neurons with acute
  or long-term reductions in sodium conductance.
\newblock {\em Journal of Neuroscience}, 25(14):3509--3520.

\bibitem[Tapia et~al., 2018]{tapia2018neurotransmitter}
Tapia, M., Baudot, P., Formisano-Tr{\'e}ziny, C., Dufour, M.~A., Temporal, S.,
  Lasserre, M., Marqu{\`e}ze-Pouey, B., Gabert, J., Kobayashi, K., and
  Goaillard, J.-M. (2018).
\newblock Neurotransmitter identity and electrophysiological phenotype are
  genetically coupled in midbrain dopaminergic neurons.
\newblock {\em Scientific reports}, 8(1):13637.

\bibitem[Taylor et~al., 2009]{taylor2009multiple}
Taylor, A.~L., Goaillard, J.-M., and Marder, E. (2009).
\newblock How multiple conductances determine electrophysiological properties
  in a multicompartment model.
\newblock {\em Journal of Neuroscience}, 29(17):5573--5586.

\bibitem[Temporal et~al., 2012]{temporal2012neuromodulation}
Temporal, S., Desai, M., Khorkova, O., Varghese, G., Dai, A., Schulz, D.~J.,
  and Golowasch, J. (2012).
\newblock Neuromodulation independently determines correlated channel
  expression and conductance levels in motor neurons of the stomatogastric
  ganglion.
\newblock {\em Journal of neurophysiology}, 107(2):718--727.

\bibitem[Tobin et~al., 2009]{tobin2009correlations}
Tobin, A.-E., Cruz-Berm{\'u}dez, N.~D., Marder, E., and Schulz, D.~J. (2009).
\newblock Correlations in ion channel mrna in rhythmically active neurons.
\newblock {\em PloS one}, 4(8):e6742.

\bibitem[Walters and Byrne, 1984]{walters1984activity}
Walters, E.~T. and Byrne, J.~H. (1984).
\newblock Activity-dependent neuromodulation: A mechanism for associative
  plasticity.
\newblock {\em Neuronal Growth and Plasticity}, 6:219.

\end{thebibliography}

\clearpage
\section*{\textbf{STAR\textbigstar METHODS}}
\subsection*{\textbf{RESOURCE AVAILABILITY}}

\subsubsection*{\textbf{Lead contact}}\mbox{}\
Further information and requests for resources should be directed to and will be fulfilled by the lead contact, Dr.\ Guillaume Drion (\href{mailto:gdrion@uliege.be}{gdrion@uliege.be}).

\subsubsection*{\textbf{Materials availability}}\mbox{}\
This study did not generate new unique reagents.

\subsubsection*{\textbf{Data and code availability}}\mbox{}\
The Julia programming language have been used in this work \citep{bezanson2017julia}. Every code and data can be found in the first author GitHub (\href{https://github.com/arthur-fyon/CORR\_2024}{https://github.com/arthur-fyon/CORR\_2024}). Numerical integration were realized using \textit{DifferentialEquations.jl}. Regression lines were computed using \textit{Polynomials.jl}. Correlations were computed using \textit{Statistics.jl}. PCA were conducted using \textit{LinearAlgebra.jl}.

\subsection*{\textbf{CONDUCTANCE-BASED MODELS}}
For all experiments, single-compartment conductance-based models were employed. These models articulate an ordinary differential equation for the membrane voltage $V$, where $N$ ion channels are characterized as nonlinear dynamic conductances, and the phospholipid bilayer is represented as a passive resistor-capacitance circuit. Mathematically, the voltage-current relationship of any conductance-based neuron model is expressed as follows:
\begin{align*}
    I_C &= C \frac{\mathrm{d}V}{\mathrm{d}t} + g_\mathrm{leak}(V-E_\mathrm{leak}) = -I_\mathrm{int} + I_\mathrm{ext}\\
    & = -\sum_{\mathrm{ion} \in \mathcal{I}} g_\mathrm{ion}(V,t) (V - E_{\mathrm{ion}}) + I_\mathrm{ext}.
\end{align*}
Here, $C$ represents the membrane capacitance, $g_\mathrm{ion}$ denotes the considered ion channel conductance and is non-negative, gated between 0 (all channels closed) and $\bar{g}_\mathrm{ion}$ (all channels opened), $E_{\mathrm{ion}}$ and $E_{\mathrm{leak}}$ are the channel reversal potentials, $\mathcal{I}$ is the index set of intrinsic ionic currents considered in the model, and $I_\mathrm{ext}$ is the current externally applied \emph{in vitro}, or the combination of synaptic currents. Each ion channel conductance is nonlinear and dynamic, represented by
\begin{equation*}
    g_\mathrm{ion}(V,t) = \bar{g}_\mathrm{ion} m_\mathrm{ion}^a(V, t) h_\mathrm{ion}^b(V, t),
\end{equation*}
where $m_\mathrm{ion}$ and $h_\mathrm{ion}$ are variables gated between 0 and 1, modeling the opening and closing gates of ion channels, respectively. Throughout this study, both the isolated crab STG neuron model \citep{liu1998model} and the adapted DA neuron model \citep{qian2014mathematical} (where SK channels had been blocked to enable bursting) were employed.

The STG model consists of seven ion channels that operate on various time scales:
\begin{itemize}
    \item fast sodium channels ($\bar{g}_\mathrm{Na}$);
    \item delayed-rectifier potassium channels ($\bar{g}_\mathrm{Kd}$);
    \item T-type calcium channels ($\bar{g}_\mathrm{CaT}$);
    \item A-type potassium channels ($\bar{g}_\mathrm{A}$);
    \item slow calcium channels ($\bar{g}_\mathrm{CaS}$);
    \item calcium controlled potassium channels ($\bar{g}_\mathrm{KCa}$);
    \item H channels ($\bar{g}_\mathrm{H}$).
\end{itemize}

The DA model consists of six ion channels that operate on various time scales:
\begin{itemize}
    \item fast sodium channels ($\bar{g}_\mathrm{Na}$);
    \item delayed-rectifier potassium channels ($\bar{g}_\mathrm{Kd}$);
    \item L-type calcium channels ($\bar{g}_\mathrm{CaL}$);
    \item N-type calcium channels ($\bar{g}_\mathrm{CaN}$);
    \item ERG channels ($\bar{g}_\mathrm{ERG}$);
    \item NMDA channels ($\bar{g}_\mathrm{NMDA}$).
\end{itemize}
Note that, owing to the multicellular nature of NMDA channels, they were excluded from this study but were still utilized for simulations with baseline values.

\subsection*{\textbf{METHOD DETAILS}}
\subsubsection*{\textbf{Random sampling sets}}\mbox{}\

Random sampling sets consist of 200 neuron models with varying maximum ion channel conductances. These sets were created by generating numerous random points in the space of maximum ion channel conductances (within specified ranges). Subsequently, the models underwent post-processing based on their firing patterns, with only those fitting the desired phenotype being retained. For the STG models, post-processing involved considerations of peak and hyperpolarized voltages, intra- and interburst frequencies, the number of spikes per burst, and burstiness (computed as in \cite{franci2018robust}). Meanwhile, the DA models were post-processed based on their peak and hyperpolarized voltages and spike frequency.

\subsubsection*{\textbf{An efficient method to build degeneracy sets that allows to remove the effect of homogeneous scaling.}}\mbox{}\

Throughout this study, a novel method for generating degenerate datasets of conductance-based models has been developed, which proves to be significantly faster than the random sampling approach (all figures were created using a dataset of 500 neurons). The methodology for a $N$-channel conductance-based model can be summarized as follows:
\begin{enumerate}
    \item 
    The leakage conductance $g_\mathrm{leak}$ is drawn from a physiological uniform distribution: $g_\mathrm{leak} \sim \mathcal{U}\left(g_\mathrm{leak\, min}, g_\mathrm{leak\, max}\right)$;
    
    \item 
    $N-3$ maximum ion channel conductances are drawn from a physiological uniform distribution that is proportional to $g_\mathrm{leak}$: $\bar{g}_\mathrm{ion} \sim g_\mathrm{leak} \cdot \frac{\mathcal{U}^{N-3}\left(\bar{g}_\mathrm{unmod\,min}, \bar{g}_\mathrm{unmod\,max}\right)}{\left(g_\mathrm{leak\, min} + g_\mathrm{leak\, max}\right)/2}$;
    
    \item 
    The 3 remaining maximum ion channel conductances are computed using the algorithm described in \citep{drion2015dynamic}.
\end{enumerate}

The normalization by $g_\mathrm{leak}$ in (ii) permits to combine the effects of homogeneous scaling and variability in conductance ratios. The subsequent sets, each targeting either homogeneous scaling or conductance ratio, were generated by using shared deterministic values for $g_\mathrm{leak}$ or for the $N-3$ maximum ion channel conductances, respectively.

The zero sensitivity directions of slow dynamical membrane properties were computed using the equations in \citep{drion2015dynamic} for the slow dynamic input conductance, where the two ion channel conductances of interest were treated as variables along this direction.

\subsubsection*{\textbf{Neuromodulation algorithm}}\mbox{}\

As a result of this newly developed method for generating degenerate neuronal sets, neuromodulation of such sets is achieved through an adaptation of the algorithm described in \citep{drion2015dynamic}, where two maximum conductances are recomputed by tuning the value of the slow dynamic input conductance while keeping the ultraslow dynamic input conductance fixed. The latest results were obtained by continuously tuning this slow dynamic input conductance value.

\end{document}